\documentclass[a4paper,12pt]{article}
\def\al{\alpha}
\def\be{\beta}
\def\ga{\gamma} \def\Ga{\Gamma}
\def\ep{\epsilon}
\def\lam{\lambda}

 \def\calH{{\cal H}} 
 \def\calK{{\cal K}} \def\calL{{\cal L}}

 \def\calT{{\cal T}}

\def\del        {  \partial  }
\def\half       {  {1\over 2}  }

\def\ie         {  {\it i.e.}      }

\def\comma          {\, ,}
\def\period         {\, .}
\def\lsim    {\lower .65ex \hbox{\ $\stackrel{<}{\sim}$\ } }
\def\gsim    {\lower .65ex \hbox{\ $\stackrel{>}{\sim}$\ } }
\def\com#1#2   { \left[#1, #2\right]} 
\def\acom#1#2  {\left\{ #1,#2\right\}}
\def\bra#1     {\langle #1 |}
\def\ket#1     {| #1 \rangle}
\def\slash#1{{\ooalign{\hfil/\hfil\crcr$#1$}}} 
\def\vecii#1#2      {  \left(\begin{array}{c}#1\\#2\end{array}\right)  }
\def\veciii#1#2#3   {  \left(\begin{array}{c}#1\\#2\\#3\end{array}
                     \right)  }
\def\veciv#1#2#3#4  {  \left(\begin{array}{c}#1\\#2\\#3\\#4
                                 \end{array}\right)  }
\def\vecfv#1#2#3#4#5 {  \left(\begin{array}{c}#1\\#2\\#3\\#4\\#5
                                 \end{array}\right)  }

\def\matrixii#1#2#3#4            {  \left(\begin{array}{cc}#1&#2\\#3&#4
                                       \end{array}\right) }
\def\matrixiii#1#2#3#4#5#6#7#8#9 {  \left(\begin{array}{ccc}#1&#2&#3\\
                                     #4&#5&#6\\#7&#8&#9\end{array}
                               \right)  }
\def\mativ#1#2#3#4               {  \left(\begin{array}{cccc}
                                       #1\\#2\\#3\\#4\end{array}\right) }
\def\matv#1#2#3#4#5              {  \left(\begin{array}{ccccc}
                                     #1\\#2\\#3\\#4\\#5\end{array}
                              \right)  }
\def\eqabegin         {  \begin{eqnarray}  }
\def\eqaend           {  \end{eqnarray}  }
\def\nn               {  \nonumber  }
\def\bracetwo#1#2     {  \left\{ \begin{array}{l} #1 \\ #2 \end{array}
                         \right.  }
\def\bracetwocases#1#2#3#4  {   \left\{ \begin{array}{ll} #1 &
                                 \qquad #2 \\
                                 #3 & \qquad #4 \end{array} \right.  }
\def\bracebegin#1     {  \left\{ \begin{array}{#1}   }
\def\braceend         {  \end{array}\right.   }
\def\parn              {  \par\noindent }

\def\parmedskip        {  \par\medskip  }

\def\parag#1           {\paragraph{#1} \mbox{ }\parmedskip\noindent}
\def\msection#1      {  \begin{center} \section{#1} \end{center}   }
\def\nsection#1      {  \let\boldface\bf \def\bf{} \section{#1}
                           \let\bf\boldface   }
\def\mnsection#1     {  \begin{center} \nsection{#1} \end{center}  }
\def\capsection#1    {  \let\boldface\bf \def\bf{\sc} \section{#1}
                           \let\bf\boldface   }
\def\mcapsection#1   {  \begin{center} \capsection{#1} \end{center} }

\newcommand{\nullify}[1]{}

\def\papertitlepage{\baselineskip 3.5ex \thispagestyle{empty}}
\def\Title#1{\baselineskip 1cm \vspace{1.5cm}\begin{center}
 {\Large\bf #1} \end{center} 
\vspace{0.5cm}}
\def\Authors#1{\begin{center} {\it #1} \end{center}}
\def\Abstract{\vspace{1.0cm}\begin{center} {\large\bf Abstract} 
           \end{center} \par\bigskip}
\def\Komabanumber#1#2#3{\hfill \begin{minipage}{4.2cm} UT-Komaba #1
              \parn #2 
              \parn #3 \end{minipage}}
\renewcommand{\thefootnote}{\fnsymbol{footnote}}
\renewenvironment{thebibliography}{\pagebreak[3]\par\vspace{0.6em}
\begin{flushleft}{\large \bf References}\end{flushleft}
\vspace{-1.0em}

\begin{enumerate}\if@twocolumn\baselineskip=0.6em\itemsep -0.2em
\else\itemsep -0.2em\fi\labelsep 0.1em}{\end{enumerate}}

\usepackage{graphicx}
\usepackage{latexsym}
\usepackage{amsmath}
\usepackage{amssymb}

\renewenvironment{thebibliography}{\pagebreak[3]\par\vspace{0.6em}
\begin{flushleft}{\large \bf References}\end{flushleft}
\vspace{-1.0em}

\begin{enumerate}\if@twocolumn\baselineskip=0.6em\itemsep -0.2em
\else\itemsep -0.2em\fi\labelsep 0.1em}{\end{enumerate} }
\def\adot{{\dot{a}}}
\def\bdot{{\dot{b}}}
\def\dotc{{\dot{c}}}
\def\dotd{{\dot{d}}}
\def\xdot{\dot{x}}

\def\aldot{{\dot{\al}}}
\def\bedot{{\dot{\be}}}

\def\Qtil{\tilde{Q}}
\def\lamtil{\tilde{\lambda}}

\def\Qhat{\hat{Q}}

\def\Gahat{\hat{\Gamma}}

\def\thhat{\hat{\theta}}

\def\Pihat{\hat{\Pi}}
\def\thtilhat{\hat{\tilde{\theta}}}
\def\What{\hat{W}}

\def\thbar{\bar{\theta}}

\def\Thbar{\bar{\Theta}}

\def\lamtil{\tilde{\lam}}
\def\Qtil{\tilde{Q}}

\def\thtil{\tilde{\theta}}

\def\ptil{\tilde{p}}
\def\ktil{\tilde{k}}

\def\Ktil{\tilde{K}}
\def\ptil{{\tilde{p}}}

\def\Dtil{\tilde{D}}

\def\th{\theta}
\def\Th{\Theta}
\def\sig{\sigma}
\def\sigp{\sigma'}

\def\pslash{\slash{p}}
\def\kslash{\slash{k}}

\def\deltassp{\delta(\sigma-\sigma')}
\def\deltapssp{\delta'(\sigma-\sigma')}

\def\Dcom#1#2{\bigl\{#1,\,#2\bigr\}_D}
\def\Dscom#1#2{\bigl\{#1,\,#2\bigr\}_{D^\ast}}
\def\Pcom#1#2{\bigl\{#1,\,#2\bigr\}_P}

\def\args{(\sigma)}
\def\argsp{(\sigma')}

\def\dels{\del_\sigma}
\def\Qone{Q^{(1)}}
\def\Qtwo{Q^{(2)}}
\def\Qthree{Q^{(3)}}
\def\npb#1{Nucl. Phys. {\bf B#1}}
\def\prd#1{Phys. Rev. {\bf D#1}}
\def\jhep#1{JHEP {\bf #1}}
\def\hepth#1{ hep-th/#1}
\def\plb#1{Phys. Lett. {\bf B#1}}

\begin{document}
\papertitlepage
\vspace*{0cm}
\Komabanumber{05-2}{hep-th/0502208} {February, 2005}
\Title{Origin of  Pure Spinor Superstring} 
\Authors{{\sc Yuri Aisaka\footnote[2]{yuri@hep1.c.u-tokyo.ac.jp} 
 and Yoichi Kazama
\footnote[3]{kazama@hep1.c.u-tokyo.ac.jp}
\\ }
\vskip 3ex
 Institute of Physics, University of Tokyo, \\
 Komaba, Meguro-ku, Tokyo 153-8902 Japan \\
  }
\vspace{1.3cm}
\baselineskip .7cm

\numberwithin{equation}{section}
\numberwithin{figure}{section}
\numberwithin{table}{section}

\parskip=0.9ex

\Abstract

The pure spinor formalism for the superstring, initiated by N.~Berkovits, 
is derived at the fully quantum level starting from a fundamental
reparametrization invariant and super-Poincar\'e invariant 
worldsheet action. It is a simple extension of the 
Green-Schwarz action with {\it doubled} spinor degrees of freedom 
with a compensating {\it local supersymmetry} on top of the conventional
$\kappa$-symmetry. Equivalence to the Green-Schwarz formalism 
is manifest from the outset. The use of free fields in the pure spinor 
formalism is justified from the first principle.
The basic idea works also for the superparticle in 11 dimensions.

\newpage
\baselineskip 3.5ex
\section{Introduction}  
\renewcommand{\thefootnote}{\arabic{footnote}}
The pure spinor (PS) formalism, initiated by 
Berkovits \cite{Berk0001}, is a remarkable construct for describing 
 the superstring. Essentially everything is controlled by 
 the nilpotent BRST-like operator $Q=\int dz \lam^\al d_\al$, where $d_\al 
 = p_\al + i\del x^m (\ga_m \th)_\al + \half (\ga^m \th)_\al 
 (\th \ga_m \del \th)$ is the spinor covariant derivative and 
$\lam^\al$ is a bosonic chiral spinor satisfying the 
  pure spinor constraints $\lam^\al \ga^m_{\al\be} \lam^\be =0$. 
 All the basic fields are postulated to be free and 
 constitute a conformal field theory 
with vanishing central charge. The cohomology of $Q$
correctly describes the 
lightcone spectrum of the superstring \cite{Berk0006}
 and appropriate vertex operators 
 and a set of rules can be given to compute the scattering amplitudes 
 in a super-Poincar\'e covariant manner \cite{Berk0001}, 
even to all loops in principle \cite{Berk0406}.
Moreover, it can be coupled to backgrounds
 including Ramond-Ramond fields in a quantizable and 
 covariant way \cite{Berk0001,Berk0009}, in distinction to 
 the conventional Ramond-Neveu-Schwarz (RNS) \cite{RNS}
 and Green-Schwarz (GS) \cite{GS}
formalisms,  where one meets difficulties. This feature makes 
the PS formalism particularly promising for deeper understanding 
of the gauge/string  correspondence \cite{Malda9711}. 
For many other developments, the reader is referred to 
  \cite{Berkothers} as well as a review \cite{Berk0209ICTP}.

Successful as it has been, there are a number of fundamental questions
  to be clarified on the PS formalism. 
One of them is the 
understanding of the quantization, especially that of the pure spinor
 $\lam^\al$. Even though the free-field postulate is powerful
 and attractive, the ghost-like field $\lam^\al$, subject to the quadratic 
 constraints, is not truly free. It is hard to imagine how such 
 a field could arise naturally in a conventional quantization process. 
Also, solving the constraints breaks manifest Lorentz covariance in the
 intermediate steps of computations. This prompted an attempt for a 
fully covariant formulation in an extended Hilbert space 
 where the pure spinor constraints 
 are removed \cite{stonybrook}. Subsequently an alternative formulation
 without pure spinor constraints, 
 which is more closely related to the original PS formalism, 
was proposed \cite{AK1}\nocite{AK2}--\cite{AK3}. 
These proposals added interesting insights and demonstrated certain 
 advantage of the enlarged field space, 
but they are yet 
to be fully developed.  Another basic question is 
the origin of the BRST-like operator 
$Q$. Since the components of $d_\al$ do not form a closed first class algebra, 
$Q$ cannot immediately be understood as a conventional BRST charge. 

Clearly, all these and other related questions have their roots in the 
lack of our knowledge of the fundamental action and its underlying symmetries 
 for the PS formalism. There exist several thought-provoking 
attempts \cite{Oda0109,Matone0206} to derive the PS formalism but 
their success, to be fair, 
has been partial. 

However, during the past year, some concrete hints 
 have been obtained which indicate that, as had been suspected, PS formalism 
 is intimately related to the GS formalism. First, in the work \cite{AK3}
 proving 
 the equivalence of the lightcone BRST treatment of the 
GS formalism to the extended version of the 
 PS formalism, it was noted that the usual 
pair of reparametrization ghosts, commonly denoted by $(b,c)$, 
can be identified with one of the five pairs of ghosts which compensate 
 for the removal of the PS constraints. 
This basic structure reappeared in 
 a more recent work \cite{BM0412}, where further important hints
 were provided. Introducing a conjugate pair of free fields
 $(\th^\al, p_\al)$ 
 into the GS formalism in the semi-lightcone gauge \cite{semiLC} described by
 the 8-component self-conjugate  $SO(8)$ chiral spinor $S_a$,
 the authors ingeniously constructed a set of 17 operators, 
 denoted in  \cite{BM0412} by  $\hat{d}_a, \hat{d}_\adot$ and 
$\Tilde{T}$,  which form a closed first class
 algebra. The corresponding BRST operator $\hat{Q}$
is readily constructed 
by introducing 16 unconstrained bosonic spinor ghosts 
$(\hat{\lam}^a, \hat{\lam}^\adot)$ and a pair of fermionic ghosts $(b,c)$.
Then it was shown that the cohomology of $\Qhat$ is
 the same as that of $Q$ of the PS formalism with the PS constraints. 
In this mechanism, the fields  $(S_a, b,c)$ play exactly the same role
 as the 5 pairs of ghosts $(b_P, c_P)_{P=1\sim 5}$ in \cite{AK1,AK3}. 

Stimulated by these developments, in particular the idea in  \cite{BM0412}
 that an extra local fermionic symmetry can lead naturally to the 
BRST charge in an extended space 
similar to the one in our formalism  \cite{AK1}\nocite{AK2}--\cite{AK3}, 
we attempted to realize this type of local fermionic symmetry from the 
 very beginning in a completely covariant fashion. 

The result of our investigation, to be described in this paper,
 is  a worldsheet action with the reparametrization and 
the super-Poincar\'e invariance, 
 from which one can {\it derive} the pure spinor formalism from the 
 first principle at the 
 fully quantum level. In Sec.~2, we shall describe our fundamental action and 
 its symmetries. The action is a simple extension of the Green-Schwarz action 
with {\it doubled} spinor degrees of freedom, $\th$ and $\thtil$, 
with a compensating {\it local supersymmetry} on top of the usual
$\kappa$-symmetry \cite{Siegel83}. 
If one gauge-fixes $\th$ to be zero 
by this extra local symmetry, one immediately recovers the conventional
 GS formalism in terms of $\thtil$. In Sec.~3, we will perform the 
  Dirac analysis of the constraints generated by our action.
 After separating out 
 the first and the second class constraints, we will impose
 the semi-lightcone  gauge for $\thtil$, without spoiling the local 
 supersymmetry and conformal symmetry. The Lorenz covariance is 
 necessarily broken for terms involving $\thtil$ but not in the sector 
 consisting of $\th$ alone. 
This procedure leads to a closed algebra of 
first class constraints under the Dirac brackets. At this stage, 
 we will encounter a grave problem that the Dirac brackets between 
the basic variables are no longer canonical, acquiring non-linear 
 modifications. Fortunately, we are able to show in Sec.~5 
 that there is a graceful way out of this apparent impasse:
 We discover that there exists a set of field-redefinitions 
such that the new basic fields become completely 
free under the Dirac bracket.  
Quantization, which is now essentially trivial, will be performed 
in Sec.~6. It is straightforward to find the 
 quantum modifications to the constraint operators so that 
 they continue to form a closed first class algebra. Remarkably, 
 this quantum algebra will be seen to be
 identical to the one engineered in \cite{BM0412}. 
The rest of the the procedure to reach the PS formalism 
was already fully described in 
 \cite{BM0412}. For the convenience of the reader, however, we shall 
 briefly reproduce  the essence of the argument 
 in a slightly more streamlined fashion. 
 This completes the derivation of the PS formalism 
 from our fundamental action. As an application of our basic idea, 
 we will briefly demonstrate in Sec.~7 that our formalism 
 works straightforwardly for the superparticle in 11 dimensions as well. 
Starting from a covariant Brink-Schwarz \cite{BS} like action one can derive 
 the PS formalism, which coincides with the one introduced in 
 \cite{Berkmembrane}, capable of describing the 11 dimensional supergravity
 in a covariant way. In Sec.~8 we briefly summarize our results and 
 indicate some interesting directions for further research.

\section{Action and its symmetries}
 The basic fields of our theory are the string coordinate $x^m$
 and two types of Majorana-Weyl spinors, 
 $\th^{A\al}$  and $\thtil^{A\al}$,  of the same chirality\footnote{We 
will describe the type IIB case in this paper. The type IIA case is similar.}. 
The vector index $m$ runs from $0$ to $9$, the spinor index 
$\al$ runs from $1$ to $16$ and 
 the index $A=1,2$ labels the two sets within each type. We will often 
distinguish them by unhatted and hatted notations, such as 
 $\th^\al \equiv \th^{1\al}$, $\thhat^\al \equiv \th^{2\al}$, 
 etc. They will eventually become left (holomorphic) and right 
(anti-holomorphic) variables. 
The worldsheet coordinate will be denoted by $\xi^i =(t,\sigma)$, $i=0,1$. 
As for the $\ga$-matrices, we use 16-dimensional $\ga^m$, which are
 real and symmetric. We will employ left derivatives throughout.  

Our fundamental action is  given by\footnote{For simplicity, we set the 
 string tension to unity until we come to the quantization.}
\begin{align}
S &= \int d^2\xi (\calL_K + \calL_{WZ})\comma  \\
\calL_K &= -\half \sqrt{-g}\, g^{ij}
 \Pi^m_i \Pi_{mj} \comma \\
\calL_{WZ} &= \ep^{ij} \Pi^m_i (W_{mj}-\What_{mj}) -\ep^{ij}
W_i^m \What_{mj}  \comma 
\end{align}
where 
\begin{align}
\Pi^m_i &\equiv \del_i x^m -\sum_Ai\del_i (\th^A \ga^m \thtil^A)
-\sum_A W^{Am}_i \comma \\
W^{Am}_i  &\equiv i\Th^A \ga^m \del_i \Th^A \comma \qquad
\Th^A \equiv \thtil^A -\th^A \period
\end{align}
It can be obtained from the conventional Green-Schwarz action 
in terms of a spinor field $\Th^A$ by the following simple procedure. 
Namely, we double the spinor degrees of freedom by setting $\Th^A=
\thtil^A -\th^A$, where $\thtil^A$ and $\th^A$ are independent fields, and 
 add interaction terms $\sum_Ai\del_i (\th^A \ga^m \thtil^A)$ in $\Pi^m_i$
 to realize important symmetries described below. 

Besides the manifest reparametrization and super-Poincar\'e invariance, 
this action possesses the following three types of fermionic symmetries. First
 the global spacetime supersymmetry is realized as 
\begin{align}
\delta \th^A &= \ep^A\comma \quad \delta \thtil^A =0 \comma 
\quad (\delta \Th^A = -\ep^A) \comma \label{globalsusyone}\\
\delta x^m &= \sum_A i\ep^A \ga^m \th^A \period \label{globalsusytwo}
\end{align}
Under this transformation  $\Pi^m_i$, and hence $\calL_K$, is invariant.
The Wess-Zumino part $\calL_{WZ}$ is also invariant 
 since the transformation
for $\Th^A$ is just as in the usual GS case. 
Secondly there is a {\it local supersymmetry} 
 defined by 
\begin{align}
\delta \th^A &= \chi^A \comma \quad \delta \thtil^A
 = \chi^A\comma \quad (\delta \Th^A =0)\comma \label{localsusy1}\\
\delta x^m&= \sum_A i\chi^A \ga^m \Th^A\comma \label{localsusy2}
\end{align}
where $\chi^A(\xi)$ is a local fermionic parameter. Since $\Th^A$ 
 is invariant,  so is $W^{Am}_i$. It is easy to check that 
 $\Pi^m_i$ is also invariant. Note that by using this symmetry
 one can gauge-fix  $\th^A$ to zero, 
upon which the action reduces precisely to the 
 conventional GS action for $\thtil^A$. Therefore the equivalence 
 of our theory to the GS formalism is obvious from the outset. 
On the other hand, as we shall see, keeping this new local 
symmetry till the end 
 will lead us naturally, though non-trivially,  to the PS formalism. 
Finally, the third fermionic symmetry present is the local 
$\kappa$-symmetry, to 
 be  described shortly. 
\section{Analysis of constraints}
In this paper, we shall analyze and quantize this system in the Hamiltonian
 formulation. The path-integral quantization will be addressed in a separate 
work. The canonical Hamiltonian density 
 is most efficiently obtained by employing the ADM parametrization
of the worldsheet metric. Namely, we parametrize the metric as 
$ ds^2 = -(N dt)^2 + \ga(d\sigma + N^1 dt)^2$, where $N$ and $N^1$ are 
 the lapse and the shift functions and $\ga$ is the spatial part of 
 the metric. Since the procedure is completely standard, we only record 
 the relevant results. The definitions of the momenta $k_\al^A$ 
and $\ktil_\al^A$ conjugate to $\th^{A\al}$ and $\thtil^{A\al}$ respectively
 produce the fermionic constraints of the form
\begin{align}
D^A_\al &= k_\al^A  + i(\kslash \thtil^A)_\al
 +i\bigl(k^m + \eta_A (\Pi^m_1 -W^{Am}_1)\bigr) (\ga_m \Th^A)_\al
=0\comma 
\\
\Dtil^A_\al &= \ktil_\al^A  -i(\kslash \theta^A)_\al
-i\bigl(k^m+ \eta_A (\Pi^m_1 -W^{Am}_1)\bigr) (\ga_m \Th^A)_\al 
=0\comma 
\end{align}
where $k^m =\del \calL/\del \xdot_m$ is the momentum conjugate 
 to $x_m$, $\kslash \equiv k^m \ga_m$, 
 and $\eta_1=-\eta_2=1$. Note that their sum, which generates 
 the local supersymmetry alluded to above, takes a very simple form. 
The Hamiltonian density is then given, 
up to the above constraints,  by 
\begin{align}
\calH &= {N\over \sqrt{\ga}} T_0 + N^1 T_1 \comma 
\end{align}
where $T_0 = \half \bigl( (k-W_1+\What_1)^2 + \Pi_1^2\bigr)$ and 
$T_1 = (k-W_1 +\What_1) \cdot \Pi_1$, 
with the notations $A^2 \equiv A^m A_m\comma A\cdot B \equiv A^m B_m$.
 Demanding consistency with 
 the vanishing of the momenta conjugate to $N$ and $N^1$, we get 
 the constraints $T_0=T_1=0$. More convenient combinations are
\begin{align}
T_+ &\equiv\half( T_0 + T_1) = {1\over 4} \Pi^m \Pi_m \comma \qquad 
T_-\equiv \half(T_0 -T_1) = {1\over 4} \Pihat^m \Pihat_m \comma 
\end{align}
where $\Pi^m$ and $\Pihat^m$ are defined by 
\begin{align}
\Pi^m & \equiv k^m -W^m_1 + \What^m_1 + \Pi^m_1 = 
 k^m +\dels x^m -\sum_A i\dels (\th^A\ga^m\thtil^A) -2W^m_1\comma  \\
\Pihat^m &\equiv k^m -W^m_1 + \What^m_1 - \Pi^m_1 = 
 k^m -\dels x^m +\sum_A i\dels (\th^A\ga^m\thtil^A) +2\What^m_1 \period
\end{align}
$T_\pm$ will be identified as   parts of the conformal generators. 
Since the coefficients $N/\sqrt{\ga}$ and 
 $N^1$ can now be taken arbitrary,  we shall choose $N/\sqrt{\ga}=1$, $N^1=0$,
 namely the conformal gauge. 
\nullify{Then $p^m$ takes the 
 form
\begin{align}
p^m &= \Pi^m_0 + W^m_1 -\What^m_1 \period
\end{align}
}

The next step is to perform the complete analysis of constraints 
 a l\`a Dirac. The Poisson brackets for the fundamental fields are 
taken as\footnote{Once we choose the sign convention
 for $\Pcom{x^m}{k^n}$, the 
minus sign for $\Pcom{\th^\al}{k_\be}$ is required for various consistency 
 of the theory, such as the closure of the constraint algebra, 
 implementation of the global supersymmetry, etc. }
\begin{align}
\Pcom{x^m\args}{k^n\argsp} &= \eta^{mn}\deltassp \comma \\
\Pcom{\th^{A\al}\args}{k^B_\be\argsp} &= -\delta^{AB} \delta^\al_\be 
\deltassp \comma \\
\mbox{rest}&=0 \period
\end{align}
In spite of the fact that the basic quantities such as 
 $\Pi^m_i$ and $T_\pm$ contain  both unhatted and hatted variables, 
the algebra of constraints turns out to neatly separate into 
the ``left'' and the 
``right''  sectors. So, to simplify the description, 
 we shall hereafter concentrate on the ``left'' sector and comment on 
 the other sector as it becomes necessary. 

In this sector, the basic Poisson brackets among $D_\al$ and $\Dtil_\al$ 
are 
\begin{align}
\Pcom{D_\al\args}{D_\be\argsp} &= 2i \ga^m_{\al\be} \Pi_m  \deltassp\comma \\
\Pcom{\Dtil_\al\args}{\Dtil_\be\argsp} &= 2i\ga^m_{\al\be} \Pi_m
 \deltassp\comma \label{DtDt}\\
\Pcom{D_\al\args}{\Dtil_\be\argsp} &= -2i \ga^m_{\al\be} \Pi_m
 \deltassp\period
\end{align}
Note that the local supersymmetry generator 
\begin{align}
\Delta_\al \equiv D_\al + \Dtil_\al
\end{align}
 has  vanishing Poisson bracket with any linear combinations
 of $D_\al$ and $\Dtil_\al$, including itself.  {}From now on, 
 we will take $\Delta_\al$ and $\Dtil_\al$ as the basic fermionic 
 constraints. 

As for $T_+$, it satisfies the Virasoro algebra of the form 
$\Pcom{T_+\args}{T_+\argsp} = 2T_+\args \deltapssp +\dels T_+\args
\deltassp $, 
where $\deltapssp \equiv \del_\sig \deltassp$. 
In fact another weakly vanishing quantity $t_+\equiv \dels\Th^\al \Dtil_\al$,
which commutes with 
$T_+$, 
forms the same Virasoro algebra. Their sum will 
 serve as the total Virasoro generator. So we have 
\begin{align}
T &\equiv T_++t_+ = {1\over 4} \Pi^m \Pi_m + \dels \Th^\al \Dtil_\al \comma \\
\Pcom{T\args}{T\argsp} &= 2T\args \deltapssp +\dels T\args
\deltassp \period
\end{align}
With respect to $T$, the basic quantities transform as conformal 
 primaries. In our scheme, a conformal primary $A_n$ of dimension $n$ 
 transforms as
\begin{align}
\Pcom{T\args}{A_n\argsp} &= nA_n\args \deltapssp + (n-1)\dels A_n\args
\deltassp \period
\end{align}
This is somewhat different from the usual form, but is equivalent to it. 
 The familiar form arises if we expand around $\sigp$ instead of $\sig$
(and regard $-T$ as the generator), 
 but the form above is computationally more convenient. One can easily 
 check that a product of conformal primaries is again a conformal 
 primary, using the formula above. $\Th^\al$ is a primary of dimension $0$, 
 while $\Pi^m, \Dtil_\al, D_\al, \dels \Th^\al$ are primaries 
 of dimension $1$. 

Now focus on the relation (\ref{DtDt}), \ie 
$\Pcom{\Dtil_\al\args}{\Dtil_\be\argsp} 
 = 2i \ga^m_{\al\be} \Pi_m  \deltassp$. As $\Pi^m \Pi_m$ is a constraint,
 we have the familiar situation that a half of  $\Dtil_\al$ is of 
 second class and the other half is of first class. To separate them 
 we shall use the light-cone decomposition. Although it breaks 
 the Lorentz covariance, it does so only for terms involving $\thtil^\al$
and does not affect $\th^\al$. This feature 
will be one of the keys for 
 producing the Lorentz covariant PS formalism in terms of  $\th^\al$
 in the end. 
We split a 10-dimensional chiral spinor $\psi^\al$ into 
 $SO(8)$ chiral and anti-chiral components as $\psi^a$ and $\psi^\adot$
 respectively and adopt the light-cone conventions such as $\ga^\pm 
 \equiv \ga^0 \pm \ga^9, \Pi^\pm = \Pi^0 \pm \Pi^9$, etc. 
The transverse components of a vector $A^m$ will be denoted as\footnote{
Although the index $i$ was previously used for the the worldsheet
 vector index, there should be no confusion.}  $A^i$ 
with $i=1\sim 8$. It will be useful to remember that 
\begin{align}
&\ga^+_{\adot\bdot} = -2\delta_{\adot\bdot} \comma \quad
 (\ga^+)^{ab} = 2\delta^{ab}\comma \quad
\ga^-_{ab} = -2\delta_{ab} \comma \quad 
  (\ga^-)^{\adot\bdot} = 2\delta^{\adot\bdot}\comma \quad 
 \mbox{rest}=0 \comma \\
&\ga^i_{a\bdot} \ga^i_{c\dotd} + \ga^i_{a \dotd}  \ga^i_{c \bdot} 
 = 2\delta_{ac} \delta_{\bdot \dotd} \period
\end{align}
The Poisson bracket for the $SO(8)$ chiral components
of $\Dtil_\al$ then reads
\begin{align}
\Pcom{\Dtil_a\args}{\Dtil_b\argsp} &= 2i \delta_{ab} \Pi^+\deltassp \period
\label{DtilDtil}
\end{align}
This  shows that $\Dtil_a$'s are the second class constraints, since, as 
is customary, we will assume that $\Pi^+$ does not vanish. 
As for $\Dtil_\adot$, we will replace it by the combination
\begin{align}
\Ktil_\adot &\equiv \Dtil_\adot -{1\over \Pi^+} \Pi^i \ga^i_{\adot b}
 \Dtil_b \period
\end{align}
This essentially generates the $\kappa$-transformations. 
The bracket of $\Ktil_\adot$ with $\Dtil_b$ is given by 
\begin{align}
\Pcom{\Ktil_\adot\args}{\Dtil_b\argsp} &= 4i \ga^i_{\adot c}\ga^i_{b \dotd}
 {\dels \Th_\dotd \Dtil_ c \over \Pi^+} \deltassp\comma 
\end{align}
which is proportional to the constraint $\Dtil_c$. The bracket of 
 $\Ktil_a$ with itself is somewhat more involved and takes the form 
\begin{align}
\Pcom{\Ktil_\adot(\sig)}{\Ktil_\bdot(\sigp)}
 &= -8i  \delta_{\adot\bdot} (\calT+\calK)\args
\deltassp + \mbox{$\Dtil$ term} \comma 
\end{align}
where ``$\Dtil$ term'' signifies a term proportional to $\Dtil_a$ and 
\begin{align}
 \calT &\equiv {T \over \Pi^+} \comma \qquad 
\calK \equiv {1\over \Pi^+}  \Ktil_\dotc \dels \Th_\dotc 
\period
\end{align}
The new operators $\calT$ and $\calK$, which are proportional to 
 the constraints, enjoy the following properties:
\begin{align}
\Pcom{\calT\args}{\calT\argsp} &= 0\comma  \\
\Pcom{\calT(\sig)}{\calK(\sigp)} 
&= {\calK\over \Pi^+}(\sig) \deltapssp\comma  \qquad 
\Pcom{\calT\args}{\Dtil_a\argsp} ={\Dtil_a \over \Pi^+}\args \deltapssp
 \comma \\
\Pcom{\calK(\sig)}{\calK(\sigp)} &= -2 {\calK \over \Pi^+}(\sig) 
\deltapssp -\del \left( {\calK \over \Pi^+}\right)(\sig) \deltassp
 + \mbox{$\Dtil$ terms}\comma  \\
\Pcom{\calK(\sig)}{\Dtil_a(\sigp)} &= -{4i \over {\Pi^+}^2}
\ga^i_{a\bdot} \ga^i_{c \dotd} \Dtil_c \dels \Th_\dotd \dels \Th_\bdot\args
\deltassp \period
\end{align}
It is also clear that their brackets with the original 
 constraints, $\Ktil_\adot$ and $T$, again close into constraints. 
This shows that they are completely of first class. Note especially that 
$\calT=T/\Pi^+$ commutes with itself because $\Pi^+$ 
 is a primary field of dimension 1 with respect to $T$. 

We may now eliminate the second class constraint $\Dtil_a$ by 
employing  the Dirac bracket $\Dcom{A\args}{B\argsp} = \Pcom{A\args}{B\argsp} 
 -\int d\sig_1 d\sig_2
 \Pcom{A\args}{\Dtil_a(\sig_1)} C^{ab}(\sig_1, \sig_2) 
\Pcom{\Dtil_b(\sig_2)}{B\argsp}$, where  
$C^{ab}$ is given, from (\ref{DtilDtil}),  by 
$C^{ab}(\sig_1, \sig_2) = (1/2i \Pi^+)\delta^{ab} \delta(\sig_1-\sig_2)$, 
and we may set $\Dtil_a=0$ strongly after computing the bracket. 
It is easy to see that the effect is simply to 
set $\Dtil_a$'s to zero in the Poisson 
 bracket relations shown above. 

To further simplify the system, we will fix the constraint $\Ktil_\adot$ (and 
hence  $\calK$) by choosing the so-called  ``semi-light-cone''
gauge \cite{semiLC} for $\thtil$. 
Namely, we will impose the condition $\ga^+\thtil =0$, or 
 equivalently $\thtil_\adot=0$. Note that this
 gauge choice does not break the global spacetime supersymmetry as defined in 
 (\ref{globalsusyone}) and (\ref{globalsusytwo}) since $\thtil$ is invariant. 
One must now make a further modification of the Dirac bracket, to be 
 denoted by $\Dscom{A}{B}$, due to this gauge-fixing. Writing $\phi_I 
 = (\thtil_\adot, \Ktil_\adot)$, it is given by 
\begin{align}
\Dscom{A\args}{B\argsp} &= \Dcom{A\args}{B\argsp} \nn\\
& \qquad  -\int d\sig_1 d\sig_2
 \Dcom{A\args}{\phi_I(\sig_1)} C^{IJ}(\sig_1, \sig_2) 
\Dcom{\phi_J(\sig_2)}{B\argsp}\comma 
\end{align}
where the matrix $C^{IJ}$, the inverse of $\Dcom{\phi_I}{\phi_J}$, 
 after setting $\phi_I=0$ takes the form 
\begin{align}
\left.C^{IJ} (\sig_1, \sig_2)\right|_{\phi_I=0}  &= 
\delta_{\adot\bdot} \delta(\sig_1-\sig_2) \matrixii{8i\calT}{-1}{-1}{0} 
\period
\end{align}
{}From this and the previous formulas, it is easy to see that 
  all the brackets between the constraints vanish 
except for $\Dscom{\Delta_\adot}{\Delta_\bedot}$, which equals 
$\Dscom{D_\adot}{D_\bedot}$ upon setting all the second class constraints to 
 zero. In this way, we obtain the following strikingly simple first class 
 algebra which governs the entire classical dynamics of the theory:
\begin{align}
\Dscom{D_\adot\args}{D_\bedot\argsp} &=
 -8i\calT\args \delta_{\adot\bdot} \deltassp\comma \\
\Dscom{D_a\args}{D_\bdot\argsp}&= \Dscom{D_a\args}{D_b\argsp} =0 \comma \\
\Dscom{\calT\args}{D_a\argsp}& =\Dscom{\calT\args}{D_\adot\argsp}=
 \Dscom{\calT\args}{\calT\argsp} =0
\period
\end{align}
Also, the explicit forms of $D_\al$ and $\calT$ are significantly simplified 
 in the semi-lightcone gauge. 

 Although we have been able to 
 simplify the structure of the theory considerably, there seems to be a large 
 price to pay: Due to the use of the Dirac bracket, brackets 
 between the fundamental variables are no longer canonical. For example, 
one finds 
 $\Dscom{x^m\args}{k^n\argsp} = \eta^{mn}\deltassp 
 + (i/2\Pi^+)(\ga^m\thtil)_a (\ga^n)_a\args \deltapssp$, 
$\Dscom{k^m \args}{k^n \argsp} = -(i/2) \dels [(1/\Pi^+)(\ga^m\Th)_a
(\ga^n\Th)_a \deltapssp]$, and so on. This is  natural 
 as the original action is highly non-linear but it is disastrous 
especially for quantization. 
\section{Free field  basis}
Remarkably, there is a gratifying solution: 
 One can find judicious redefinitions 
 of the fundamental fields in such a way that the Dirac brackets among the 
 new variables are exactly canonical. In other words, there is a basis in which 
 all the fields are free. Below we summarize our results, now including
 the ``right'' as well as the ``left'' sector.
 The string coordinate $x^m$ is unchanged. 
The new momenta $p^m$, $p^A_a$ and $p^A_\adot$ are 
 given by 
\begin{align}
p^m &\equiv k^m -i\dels(\thtil \ga^m \th) + i\dels(\thtilhat \ga^m \thhat) 
\comma \label{defpm}
\\
p^A_a &\equiv k^A_a - i\eta_A (\dels x^+ -i  \th^A\ga^+ \dels\th^A)
\thtil_a^A \noindent\\
&\qquad + \eta_A 
 \bigl[
   2(\ga^i\dels\th^A)_a \thtil^A \ga^i \th^A
   + (\ga^i\th^A)_a \dels (\thtil^A \ga^i \th^A)
 \bigr] \comma 
\\
p^A_\adot &\equiv k^A_\adot +i\eta_A (\ga^m \th^A)_\adot 
\bigl[ -2i \thtil^A \ga_m \dels \th^A + i\thtil^A\ga_m \dels \thtil^A 
 -i\dels (\thtil^A \ga_m \th^A) \bigr] \nn\\
& \qquad -i\eta_A (\ga^i \thtil^A)_\adot 
 \bigl[ \dels x^i - 3i  \th^A\ga^i \dels \th^A +2i \th^A \ga^i \dels \thtil^A
 + i \dels (\thtil^{A'} \ga^i \th^{A'}) \bigr] \comma 
\end{align}
where the index $A'$ for the last term of $p^A_\adot$ signifies 
the one opposite to $A$, namely $A'=1(2)$ if $A=2(1)$. As for 
$\thtil^A$, we actually need to regard the combinations
\begin{align}
S_a &= \sqrt{2\Pi^+}\, \thtil_a \comma \quad \hat{S}_a = 
 \sqrt{2\Pihat^+}\, \thtilhat_a \comma \label{defSa}
\end{align}
as our fundamental fields. With these redefinitions it is
 straightforward to verify the following canonical Dirac bracket relations:
\begin{align}
\Dscom{x^m\args}{p^n\argsp} &= \eta^{mn} \deltassp \comma \qquad 
\Dscom{\th^{A\al}\args}{p^B_\be\argsp} = -\delta^{AB} 
\delta^\al_\be \deltassp \comma \\
\Dscom{S^A_a\args}{S^B_b\argsp} &= i\delta^{AB}\delta_{ab} 
\deltassp \comma \qquad 
\mbox{rest} = 0 \period
\end{align}
Another  non-trivial and 
satisfying feature of the above redefinitions is that in terms of 
 the new fields {\it complete separation}
 of the left and right sectors takes place:
 Not only does the algebra of constraints 
close separately in each sector (as has already been the case even before the redefinitions), the constraints in the left (right) sector 
 are now expressed solely in terms  of the left- (right-) variables. 
For instance, the form of  $\Pi^m$, which is a building block of 
 $\calT$, changes as
\begin{align}
\Pi^m &= k^m + \eta[\dels x^m -2i \th\ga^m\dels \th  - 2i  
\thtil\ga^m \dels \thtil +4i \thtil\ga^m\dels\th]  - i \dels(\thtil \ga^m \th) +i\dels(\thtilhat \ga^m \thhat)  \nn\\
&= p^m + \eta[\dels x^m -2i \th\ga^m\dels \th  - 2i  
\thtil\ga^m \dels \thtil +4i \thtil\ga^m\dels\th]\comma \label{newPim}
\end{align}
where we used the ``left-favored'' notation that for the left sector $\eta=+1$ 
and the fields are as shown, while for the right sector (\ie for $\Pihat^m$)
 $\eta=-1$ and we should put hats on $\th$ and $\thtil$ 
for the part within the parenthesis $[\quad ]$. 
This convenient notation will be used for the rest of this paper. 
Evidently the shift from $k^m$ to $p^m$ removes the variables 
 of the ``wrong sector''. 

Now let us display the explicit form of the constraints 
 in terms of the canonical free fields,  obtained by 
 using (\ref{defpm}) $\sim$ (\ref{defSa}). Employing the 
 ``left-favored'' notation introduced above, the results are
\begin{align}
D_a &= d_a  + i\sqrt{2\Pi^+}\, S_a \comma \label{Daclas}\\
D_\adot &= d_\adot + i \sqrt{{2 \over \Pi^+}} 
\Pi^i (\ga^i S)_\adot +{2\eta \over \Pi^+} (\ga^i S)_\adot 
(S \ga^i \dels \th) \comma
\\
\calT &= {1\over 4} {\Pi^m \Pi_m \over \Pi^+}
\period \label{calTclas}
\end{align}
Here 
 $\Pi^m$ is given in (\ref{newPim}) and 
 $d_a$ and $d_\adot$ are the $SO(8)$ components of 
 the covariant spinor $d_\al$ defined by 
\begin{align}
d_\al  &\equiv  p_\al  -i(\ga^m \th)_\al ( p_m + \eta \del x_m) 
 -\eta(\ga^m\th)_\al (\th \ga_m \del \th) \period
\end{align}
Remarkably, the 17 first class constraints 
 (\ref{Daclas}) $\sim$ (\ref{calTclas}) 
 will be seen to be identical, upon quantization, 
 to the ones constructed in \cite{BM0412}. 
\section{Quantization and derivation of PS formalism}
Now that we have expressed all the constraints in terms of free fields, 
 the quantization of the basic variables is essentially trivial. First, 
 replacement of the Dirac brackets by the quantum brackets yields
$\com{p^m\args}{x^n\argsp} = -i\eta^{mn}\deltassp$, 
$\acom{p^A_\al\args}{\th^{A\be}\argsp} = -i\delta_\al^\be \deltassp$ 
 and $\acom{S_a\args}{S_b\argsp} = -\delta_{ab}$. Next 
 we translate them to the OPE's in the Euclidean formulation. 
To obtain the standard normalization, we reinstate the string
 tension $T=1/(2\pi \al')= 1/(4\pi)$ with the choice $\al' =2$, Euclideanize, 
 make a conformal transformation to the plane coordinate, and make 
the identification and redefinition\footnote{Appropriate removal of the 
 factor of $z^h$ produced for an operator of dimension $h$ by 
 the conformal transformation is implicitly understood.}
of the form 
 (focussing for simplicity on the left sector)
$p^m + (1/4\pi) \dels x^m \rightarrow  (i/2\pi)\del x^{m}$,
$p_\al \rightarrow p_\al/(2\pi i)$, and $S_a \rightarrow -iS_a/\sqrt{2\pi}$, 
 where $\del \equiv \del_z$. 
 Then 
the OPE's for the basic variables become
\begin{align}
x^m(z) x^n(w) &= -\eta^{mn} \ln (z-w) \comma \quad 
p_\al (z) \th^\be(w) = {\delta_\al^\be \over z-w} \comma \quad 
S_a (z) S_b(w) = {\delta_{ab} \over z-w} \period
\end{align}
Correspondingly, it is convenient to make the following replacements
\begin{align}
D_\al &\rightarrow {1\over 2\pi i} D_\al\comma \quad d_\al 
\rightarrow {1\over 2\pi i} d_\al\comma \quad 
\Pi^m \rightarrow {1\over 2\pi} \Pi^m \comma 
\end{align}
and afterwards rescale $\calT$ so that 
 $\calT  \equiv (1/2) \Pi^m\Pi_m /\Pi^+$. 
Further, to facilitate the comparison with the result of \cite{BM0412}, 
 we will make explicit the dependence on $S_a$ by introducing 
 a quantity $\pi^m$ defined by 
\begin{align}
\pi^m &\equiv  i \del x^m + \th \ga^m \del \th \period
\end{align}
This is  nothing but the $S_a$-independent part of $\Pi^m$, which 
 after the Euclideanization and rescaling reads
\begin{align}
\Pi^m &= \pi^m + i \left( {i \over 2\pi^+}S\ga^m \del S
 -\sqrt{{2\over \pi^+}} S \ga^m \del \th\right) \period
\end{align}
Note that $\Pi^+ =\pi^+$ holds. 
Then, the redefined constraints in terms of the quantized fields 
take the form 
\begin{align}
d_\al &=  p_\al + i\del x^m (\ga_m \th)_\al + \half (\ga^m \th)_\al 
 (\th \ga_m \del \th) \comma \\
D_a &= d_a + i\sqrt{2\pi^+} S_a \comma \\
D_\adot &= d_\adot + i \sqrt{{2\over \pi^+}} \pi^i 
(\ga^i S)_\adot -{1\over \pi^+} (\ga^i S)_\adot (S\ga^i \del \th) \comma \\
\calT &= \half {\pi^m \pi_m \over \pi^+} -{1 \over 2\pi^+} S_c \del S_c
 + i \sqrt{{2\over \pi^+}} S_c \del \th_c +i {\sqrt{2}
\over (\pi^+)^{3/2}} \pi^i (S\ga^i \del\th) 
 -{1\over (\pi^+)^2} (S\ga^i \del \th)^2 \period
\end{align}

These are as yet 
the naive classical expressions written in terms of quantum fields. 
As it commonly happens, we need to add  a few improvement terms
in order to realize the local symmetry quantum mechanically. 
The necessary modifications must be related to the normal-ordering 
 ambiguities and  should cancel the double and higher poles 
arising from the multiple contractions which are absent in the classical
 computations. As for $D_\al$ there is only one term which requires 
 normal ordering, namely the term $ (-1/\pi^+) (\ga^i S)_\adot 
 (S \ga^i \del \th)$ in $D_\adot$. Therefore we expect that the terms 
 we may need are of the type $\del^2 \th_\adot/\pi^+$ and $\del\th_\adot
\del (1/\pi^+)$. Indeed by adjusting their coefficients 
 properly, the double and the triple poles in $D_\adot(z)D_\bdot(w)$ 
 can be cancelled exactly. In this way the complete quantum constraints 
 are obtained by the modifications
\begin{align}
D_a &\rightarrow D_a  \comma  \\
D_\adot &\rightarrow D_\adot + {4\del^2 \th_\adot \over \pi^+} 
 -{2\del \pi^+ \del \th_\adot \over (\pi^+)^2} \comma \\
\calT &\rightarrow   \calT + {4 \del^2 \th_\dotc \del \th_\dotc
 \over (\pi^+)^2} -\half {\del^2 \ln \pi^+ \over \pi^+}  \period 
\end{align}
Now they close under the OPE as
\begin{align}
D_\adot(z) D_\bdot(w) &= {-4\delta_{\adot\bdot} \calT(w) \over z-w} 
\comma \qquad \mbox{other OPE's $=$ regular} \period
\end{align}
Up to some difference in conventions, this result agrees precisely 
 with the one constructed in \cite{BM0412} by adding 
 free fields $(p_\al, \th^\al)$ to the GS formalism in the semi-lightcone
 gauge\footnote{The idea of adding extra free spinors to construct a 
 first class algebra appeared earlier for the $d=10$ \cite{Berk0209ICTP}
 and $d=11$ superparticle \cite{Anguelovaetal}.}. 
In the present formalism, we have been able to derive it 
 rather straightforwardly from the fundamental action
 together with the  justification  of the use of free fields, which has 
 hitherto been a postulate. 

The rest of the procedure to get to 
 the PS formalism was already fully explained in \cite{BM0412}. 
However, for completeness and for the convenience of the reader, 
 we shall briefly reproduce the argument (restricting to the holomorphic 
 sector) in a slightly more streamlined  fashion below. 

 First from the simple 
structure of the constraint algebra, one can immediately construct, 
 in a completely conventional way, 
 the nilpotent BRST charge $\Qtil$ in the form
\begin{align}
\Qtil &=\int {dz \over 2\pi i} 
 \left( \lamtil^\al D_\al +  \calT c  - (\lamtil \ga^+ \lamtil) b\right)
  \comma 
\end{align}
where $\lamtil^\al$ is an {\it unconstrained}
 bosonic spinor ghost and $(b,c)$ is a 
canonical pair of fermionic ghosts satisfying $b(z)c(w) = 1/(z-w)$. 
At this stage the theory is defined in an extended Hilbert space 
similarly to the formulation in \cite{AK1}. Due to this feature, 
 one can construct the ``$B$-ghost'' and express the energy-momentum 
 tensor as 
\begin{align}
B &= b \pi^+ - \tilde{\omega}_\al
 \del \th^\al \comma \\
T &= \bigl\{\Qtil,\,B\bigr\} = \half \pi^m \pi_m -d_\al 
 \del \th^\al - \tilde{\omega}_\al \del \lamtil^\al 
 -b \del c -\half \del^2 \ln \pi^+ \comma 
\end{align}
where $\tilde{\omega}_\al$ is the field conjugate to $\lamtil^\al$ satisfying 
 $\lamtil^\al(z) \tilde{\omega}_\be(w) = \delta^\al_\be/(z-w)$. 

The next step is to show that the cohomology of $\Qtil$ is the same as 
that of 
\begin{align}
\Qone= \int {dz\over 2\pi i}  \left( \lamtil_a D_a 
 + \lam_\adot D_\adot \right)\comma \quad \lam_\adot \lam_\adot= 0 \comma 
\end{align}
 obtained from $\Qtil$ by dropping the terms containing $(b,c)$ and 
imposing a constraint  $\lam_\adot \lam_\adot =0$ or $\lam \ga^+\lam =0$. 
(We remove tilde to indicate that it is constrained.) 
Note that this is one of the five independent 
 constraints expressed by the pure spinor conditions $\lam \ga^m \lam =0$.
One way to do this is to employ the logic of the homological 
 perturbation theory \cite{HenTb}, which is essentially the somewhat 
 lengthy analysis  presented in  \cite{BM0412}. A more direct method 
 is to connect $\Qtil$ and $\Qone$ 
 by the following similarity transformation, 
 which can be easily checked:
\begin{align}
&e^X \Qtil e^{-X} = \delta_b + \Qone   \comma \\
& \delta_b = 2\int {dz \over 2\pi i} 
\lamtil_\bdot \lamtil_\bdot b \comma \quad 
X = \int {dz \over 8\pi i}  c (l_\adot D_\adot) \period
\end{align}
Here $l_\adot$ is an auxiliary $SO(8)$ anti-chiral spinor with the 
 property $\lamtil_\adot l_\adot =1$, $l_\adot l_\adot =0$ and 
 $\lam_\adot$ in $\Qone$ is given by $\lam_\adot =
\lamtil_\adot -\half (\lamtil_\bdot \lamtil_\bdot) l_\adot$. 
This indeed satisfies $\lam_\adot \lam_\adot =0$ and we may hereafter 
 forget about $l_\adot$ as it appears 
 only in $\lam_\adot$. Since 
 $\delta_b$ is completely independent of $\Qone$ and  has 
 a trivial cohomology, we may drop it to obtain $\Qone$.

The final process is 
 to cohomologically decouple $S_a$ together with four more degrees of freedom of 
$\lamtil^\al$ by a judicious similarity transformation.  
To this end, define
the following projection operators
 $P^1$ and $P^2$  in the $SO(8)$-chiral 
 space:
\begin{align}
\delta_{ab} &= P^1_{ab} + P^2_{ab} \comma \qquad 
 P^1_{ab} \equiv  \half (\ga^i \lam)_a (\ga^i r)_b \equiv  P^2_{ba} \comma 
\end{align}
where again an anti-chiral 
 spinor $r_\adot$ with the properties $\lamtil_\adot r_\adot= 1$, 
 $r_\adot r_\adot=0$ has been introduced\footnote{%
As remarked in \cite{BM0412}, dimension $0$ operator 
 $\theta_{\dot{a}}r_{\dot{a}}$ containing $r_{\dot{a}}$
 should be excluded from the Hilbert space to avoid 
 the triviality of the cohomology.}.
Using these projection operators, 
 one can decompose the self-conjugate field $S_a$ into a ``conjugate pair''
 $(S_a^1,S_a^2)$  as $S_a^I = P^I_{ab} S_b, (I=1,2)$. They satisfy 
 the OPE's $S_a^1(z) S_b^1(w) = S_a^2(z) S_b^2(w) =\text{(regular)}$, 
$S_a^1(z) S_b^2(w) = P^1_{ab}/(z-w)$ and $S_a^2(z) S_b^1(w) = P^2_{ab}/(z-w)$. 
Similarly, $\lamtil_a$ is decomposed into 
$\lamtil_a = \lam_a^1+ \lam_a^2$, where 
$\lam_a^I = P^I_{ab}\lamtil_b$. It is important to note that $\lam_a^1$ 
is easily checked to satisfy 
\begin{align}
\lam_a^1 \ga^i_{a\bdot} \lam_\bdot =0 \comma \label{lamoneeq}
\end{align}
which are the remaining four independent equations contained in 
 the pure spinor constraints. 

The similarity transformation is best performed in two 
steps\footnote{We reverse the order of the two transformations compared to \cite{BM0412} for added clarity.}. First, 
we make a transformation $\Qtwo= e^Y \Qone e^{-Y}$ with 
\begin{align}
Y  &= -\half \int{dz \over 2\pi i} S_a^1S_a^2 \ln \pi^+\period
\end{align}
The main effect of this transformation is the replacement
\begin{align}
S_a^1 \rightarrow {S_a^1 \over \sqrt{\pi^+}} \comma 
\qquad S_a^2 \rightarrow \sqrt{\pi^+}\, S_a^2\comma 
\end{align}
which shifts the conformal weight of $(S_a^1, S_a^2)$ 
 from $(1/2, 1/2)$ to $(1,0)$, the latter being more natural for a conjugate 
 pair. Including the remaining effects, $\Qtwo$ becomes
\begin{align}
\Qtwo &= \delta + Q + d\comma \\
\delta &= \sqrt{2}\, i \lam^2_a S_a^1 \comma \qquad 
 Q = \lam^1_a d_a + \lam_\adot d_\adot  \comma \\
d&=    {4 \lam_\adot \del^2\th_\adot \over \pi^+} 
 - {4 \del \pi^+ \lam_\adot \del \th_\adot \over (\pi^+)^2}
 + \lam_a^2 d_a \nn\\
& +\sqrt{2}\, i (\pi^+ \lam_a^1 S_a^2 + \pi^i (\lam \ga^i S^2)) 
 -(\lam \ga^iS^2) (S^2 \ga^i \del \th) \comma 
\end{align}
where the first two terms in $d$ come from truly quantum contributions. 
Note that if we assign the degrees $\deg (S_a^1, S_a^2) =(-1,+1)$, 
$d$ includes all the terms of positive degree while $\delta$ and $Q$ 
 carry degrees $-1$ and $0$ respectively. With this grading structure 
 in mind,  it is an easy matter 
 to find a similarity transformation which removes $d$ entirely:
\begin{align}
\Qthree &= e^Z \Qtwo e^{-Z} = \delta + Q \comma \\
Z &= -{d_a S^2_a \over \sqrt{2} \, i} 
 +{4(\del \th_\adot\lam_\adot) (\del\th_\bdot r_\bdot) \over \pi^+} \period
\end{align}
Now since $\delta$ is completely independent of $Q$ and its cohomology is 
 easily seen to be trivial, we may drop $\delta$ as well. Finally, 
by renaming $\lam^1_a \rightarrow \lam_a$ and denoting $\lam^\al = (\lam_a, 
\lam_\adot)$, we reach the simple BRST operator $Q$ of the 
 PS formalism:
\begin{align}
Q &= \int {dz \over 2\pi i} \lam^\al d_\al\comma \qquad 
 \lam \ga^m \lam =0 \comma \\
d_\al &= p_\al + i\del x^m (\ga_m \th)_\al + \half (\ga^m \th)_\al 
 (\th \ga_m \del \th) \period
\end{align} 
\section{Application to superparticle in 11 dimensions}
Evidently, our formalism described above for the superstring 
contains, as its zero mode sector, the case of a superparticle 
 in 10 dimensions. It is in fact much 
 simpler than for the superstring, 
because the bulk of the non-trivial features of the string case, such as 
 the necessity of the redefinitions to get free fields etc., are due 
 to  expressions involving $\sigma$-derivatives, and  they are 
 absent for a  particle. 

As we will now briefly show, our basic idea works almost verbatim 
for a superparticle in 11 dimensions as well: Starting from a covariant 
Brink-Schwarz type action one can straightforwardly derive the 
PS formalism\footnote{Strictly speaking, the notion of 
the pure spinor in 11 dimensions in the sense of Cartan \cite{Cartan} requires 
additional conditions, but we will continue to refer to it as PS 
 formalism.} for it, which coincides with the one 
 introduced in  \cite{Berkmembrane}. 

Let us first summarize the conventions and properties of the
  $\Ga$-matrices and spinors to be used.
  $32 \times 32$ real $\Ga$-matrices will be
 denoted by $\Ga^M_{AB}$, $(M=0 \sim 10;\; A,B=1\sim 32)$,
 the charge conjugation matrix $C=\Ga^0$ 
 has the property $C^T = -C$ and $C\Ga^M$ and $\Ga^MC$ are symmetric. 
The lightcone decomposition of $\Ga^M$ is taken as 
 $\Ga^\pm = \Ga^0 \pm \Ga^{10}$, $\Ga^i$ ($i=1\sim 9$). 
The lightcone chirality operator 
(\ie $SO(1,1)$ boost  charge) is defined as $\Gahat \equiv \Ga^0\Ga^{10}$ 
and a 32-component 
spinor $\chi_A$  will be decomposed according to the eigenvalues 
 of $\Gahat$ as 
$\chi_A =(\chi_\al, \chi_\aldot)$, $(\al, \aldot =1\sim 16)$, with 
 $\Gahat_{\al\be}\chi_\be = +\chi_\al$, $\Gahat_{\aldot\bedot} \chi_\bedot
 = -\chi_\aldot$. Note that this decomposition is in parallel with
 the $SO(8)$ decomposition for the 10-dimensional case and differs from 
 the one with respect to the 10-dimensional chirality operator $\Ga^{10}$. 

We start from the covariant action of the form
\begin{align}
S &= \int dt L \comma \qquad L = {1\over 2e} \Pi^M \Pi_M \comma 
\end{align}
where $e$ is the einbein and $\Pi^M$ is given by
\begin{align}
\Pi^M &= \xdot^M -i\del_t(\thbar \Ga^M \thtil) + i\dot{\Thbar} \Ga^M \Th
\comma \qquad \Th \equiv \thtil -\th\comma
\end{align}
where $\thbar$ denotes the usual Dirac conjugate $\th C$. 
This action is invariant under the reparametrization, the super-Poincar\'e 
transformation, and the following three fermionic transformations:
 Namely, the global supersymmetry transformation
\begin{align}
\delta \th_A &= \ep_A\comma \quad \delta \thtil_A =0\comma 
\quad \delta x^M = i\bar{\ep}\Ga^M \th \comma 
\end{align}
the local supersymmetry transformation
\begin{align}
\delta \th_A &= \delta \thtil_A = \chi_A \comma 
\quad \delta x^M = i\bar{\chi} \Ga^M \Th \comma 
\end{align}
and the $\kappa$-transformation, to be described below. 

After the standard Hamiltonian analysis and choosing the gauge $e=1$, 
 one finds that the total Hamiltonian consists of  arbitrary linear 
 combination of the constraints of the form
\begin{align}
D_A &= p_A -i(C\pslash (\theta -2\thtil))_A=0 \comma \\
\Dtil_A &= \ptil_A -i(C\pslash\thtil)_A =0\comma \\
T &=\half p^2 =0\comma 
\end{align}
where $p^M$, $p_A$ and $\ptil_A$ are the momenta conjugate to 
$x_M, \th_A$ and $\thtil_A$  respectively and $\pslash \equiv p_M \Ga^M$. 
With the basic Poisson brackets $\Pcom{x^M}{p^N}
=\eta^{MN}, \Pcom{\th_A}{p_B} = -\delta_{AB}$ and $\Pcom{\thtil_A}{\ptil_B}
 = -\delta_{AB}$, they satisfy the  algebra
\begin{align}
\Pcom{D_A}{D_B} &= \{\Dtil_A, \Dtil_B\} =2i(C\pslash)_{AB} 
\comma \qquad \{D_A, \Dtil_B\}_P = -2i (C\pslash)_{AB} \comma \\
\Pcom{T}{T} &= \Pcom{T}{D_A} = \{T, \Dtil_A\}_P =0 \period
\end{align}
Again the combination $\Delta_A \equiv D_A + \Dtil_A$ commutes with 
 all the constraints. 

To dissociate the second class part 
from the  the first class part, 
 we invoke the decomposition with respect to the lightcone chirality 
 and employ the $\kappa$-symmetry generator $\Ktil_\aldot$ in place of 
 $\Dtil_\aldot$:
\begin{align}
\Ktil_\aldot &\equiv \Dtil_\aldot -{1\over p^+} p^i
(C\Ga^i)_{\aldot\be}  \Dtil_\be \period
\end{align}
Then we get the algebra
\begin{align}
\Pcom{D_\al}{D_\be} &= 2i p^+ \delta_{\al\be} \comma \quad 
\Pcom{\Ktil_\aldot}{D_\be} = 0 \comma \\
\Pcom{\Ktil_\aldot}{\Ktil_\bedot} &= -4i {T \over p^+} 
\delta_{\aldot\bedot} \comma 
\end{align}
showing that $\Dtil_\al$ is of second class and $\Ktil_\aldot$ is of 
first class.
We now take the semi-lightcone gauge $\thtil_\aldot =0$ to render 
 the pair $(\Ktil_\adot, \thtil_\aldot)$ second class 
 and introduce the total Dirac bracket $\Dscom{\star}{\star}$ with respect to all the 
 second class constraints including $\Dtil_\al$. Then, the variable 
 $\thtil_\al$, with a rescaling, becomes self-conjugate as 
\begin{align}
\Dscom{S_\al}{S_\be} &= i\delta_{\al\be} \comma \qquad S_\al \equiv 
 \sqrt{2p^+} \thtil_\al \comma 
\end{align}
and we are left with a completely first class constraint algebra:
\begin{align}
\Dscom{D_A}{D_B} &= 2i {T \over p^+} (C\Ga^+)_{AB} \comma \quad 
 \Dscom{T}{D_A} = 0\comma \quad \Dscom{T}{T} =0 \period
\end{align}

As said before, for the particle case all the basic variables 
are already free and we can readily quantize the theory in the standard 
 way. For convenience we make  rescalings $S_\al \rightarrow 
 -iS_\al$, $p_A \rightarrow -ip_A$ 
 so that the quantized variables obey simpler (anti)commutation relations:
\begin{align}
\com{p^m}{x^n} &= {1\over i} \eta^{mn}\comma \quad 
\acom{p_A}{\th_B} = \delta_{AB} \comma \quad \acom{S_\al}{S_\be} 
 = \delta_{\al\be} \period
\end{align}
We also make a redefinition $D_A \rightarrow -iD_A$. Then the quantum 
 constraints take the form 

\begin{align}
D_A &= d_A + \delta_A \comma \qquad T = \half p^2 \comma \\ 
d_A &= p_A + (C\pslash)_{AB} \th_B \comma\quad
 \delta_\al = i\sqrt{2p^+} S_\al \comma \quad 
\delta_\aldot = i \sqrt{{2\over p^+}} 
p^i (C\Ga^i)_{\aldot \be} S_\be \comma 
\end{align}
and they form the algebra 
\begin{align}
\acom{D_\aldot}{D_\bedot} &= -4{T \over p^+} \delta_{\aldot\bedot} \comma 
\quad \acom{D_\al}{D_\be} = \acom{D_\al}{D_\bedot} = \com{T}{D_A} = 
 \com{T}{T} =0 \period 
\end{align}
The corresponding BRST operator is 
\begin{align}
\Qhat &= \lam_A D_A + \lam_\aldot \lam_\aldot  b + {2T \over p^+} c\comma
\end{align}
where $\lam_A$ and  $(b,c)$ respectively
 are the bosonic and fermionic ghosts.  
This is precisely of the same structure as the one for the 
 10 dimensional superparticle discussed in  \cite{BM0412} and 
 derived in this paper as a part of the superstring. Moreover, 
 the rest of the procedure to decouple $S_\al$ together with a part of 
 $\lam_A$, elaborated in  \cite{BM0412}, 
 leading to the PS formalism goes through 
 verbatim by replacing the $SO(8)$ $\ga$-matrices $\ga^i$, $i=1\sim 8$ 
with the $SO(9)$ $\Ga$-matrices $C\Ga^i$, $i=1\sim 9$. The final result 
 is the BRST operator $Q=\lam^A d_A$, with the PS conditions
 $\lam \Ga^M C \lam =0$ (or $\lam C\Ga^M \lam=0$ if we redefine $C\lam 
 \rightarrow \lam$) for $M=0 \sim 10$. This was shown in 
\cite{Cederwalletal,Berkmembrane} to be 
 the correct conditions to reproduce the spectrum of the supergravity 
 in 11 dimensions as the cohomology of $Q$.

\section{Summary and discussions}
In this paper, we have constructed a reparametrization and super-Poicar\'e
 invariant worldsheet action from which one can derive the pure spinor 
 formalism in a logically complete manner. The basic idea was to write 
 the Green-Schwarz spinor field $\Th$ as a difference of 
two independent fields $\thtil -\th$ 
and at the same time introduce  appropriate 
 interactions between them so that an extra compensating local supersymmetry 
is realized. By fixing the gauge for $\thtil$ using $\kappa$-symmetry 
while untouching $\th$ and retaining the new local supersymmetry,
 one is lead to a simple closed system of first 
 class constraints. Moreover, we found  highly non-trivial 
 redefinitions of fields, 
 under which all the basic variables become canonically free. 
This allowed us to quantize the theory in a straightforward manner with 
 slight quantum modifications for the form of the constraints. 
Remarkably this set of quantized constraints agreed precisely with 
 those engineered in  \cite{BM0412}. Then as demonstrated in  \cite{BM0412}
 one can immediately construct the BRST operator 
and show  that its cohomology is  equivalent to that in the PS formalism. 
We have also shown that our idea works equally well for the superparticle 
 in 11 dimensions, with the emergence of the correct constraints for 
 the bosonic spinor ghosts to describe the 11 dimensional supergravity.

There are many interesting further investigations one would like to perform
 based on the present formalism. Let us briefly discuss some of them below. 
\begin{itemize}
\item One obvious problem is the path-integral reformulation of our 
idea. In particular,
 now that we can start form a fundamental action, it should be 
 possible to derive the appropriate measure \cite{Berk0001, Berk0406}
 and gain deeper understanding 
 from the first principle. 
\item Another intriguing project is the application to the supermembrane. 
The success for the superparticle in 11 dimensions is an encouraging sign but 
the full-fledged investigation for the supermembrane is expected to be 
highly non-trivial. In any case, such a study  would no doubt shed 
 a new light on the structure  of the supermembrane dynamics.

\item Although we believe that our fundamental action is the minimal one 
containing all the necessary ingredients, it need not be unique.
 There might be some advantage to embed it in a larger framework with 
further (local) symmetries, for instance a doubly supersymmetric 
 formulation, so that added freedom to  manipulate the gauge choice 
may produce interesting variants. 
\end{itemize}

Some of these and related problems are currently 
under investigation and we hope 
 to report on the results in future communications. 
\par\bigskip\noindent
{\large\bf Acknowledgment}\par\smallskip\noindent
The research of Y.K. is supported in part by the 
 Grant-in-Aid for Scientific Research (B) 
No.~12440060 and (C) No.~15540256 from the Japan 
 Ministry of Education,  Science and Culture. 
\newpage
\nullify{
\appendix
\setcounter{equation}{0}
\renewcommand{\theequation}{A.\arabic{equation}}
\section*{Appendix A: \ \
Conventions and Useful Formulas}
}


\begin{thebibliography}
\bibitem{Berk0001} 
 N.~Berkovits, \jhep{0004} (2000) 018, \hepth{0001035}. 
%
\bibitem{Berk0006} 
 N.~Berkovits, \jhep{0009} (2000) 046, \hepth{0006003}.
%
\bibitem{Berk0406} 
 N.~Berkovits, \jhep{0409} (2004) 047, \hepth{0406055};
 \hepth{0410079}.
%
\bibitem{Berk0009}
 N.~Berkovits and O.~Chandia, \npb{596} (2001) 185, \hepth{0009168}.
%
\bibitem{RNS}
P.~Ramond, \prd{3} (1971) 2415; \\ 
A.~Neveu and J.~H.~Schwarz, \npb{31} (1971) 86. 
\bibitem{GS}
M.~B.~Green adn J.~H.~Schwarz, 
 \npb{181} (1981) 502;
 \plb{136} (1984) 367;
 \npb{243} (1984) 285.
\bibitem{Malda9711}
 J.~M.~Maldacena,
 Adv.\ Theor.\ Math.\ Phys.\  {\bf 2} (1998) 231,
 \hepth{9711200}.
%
\bibitem{Berkothers}
  N.~Berkovits and B.~C.~Vallilo, \jhep{0007} (2000) 015, \hepth{0004171};
%
 N.~Berkovits, 
  Int.\ J.\ Mod.\ Phys.\ A {\bf 16} (2001) 801, \hepth{0008145};
%
 \jhep{0108} (2001) 026, \hepth{0104247};
%
 \jhep{0109} (2001) 016, \hepth{0105050};
%
  N.~Berkovits and O.~Chandia, \plb{514} (2001) 394,  \hepth{0105149};
%
  N.~Berkovits and P.~S.~Howe, \npb{635} (2002) 75, \hepth{0112160};
%
  N.~Berkovits, 
  \jhep{0204} (2002) 037, hep-th/0203248;
%
  N.~Berkovits and O.~Chandia, \jhep{0208} (2002) 040, \hepth{0204121};
%
  N.~Berkovits and V.~Pershin, \jhep{0301} (2003) 023, \hepth{0205154};
%
G.~Trivedi, Mod.\ Phys.\ Lett.\ A {\bf 17} (2002) 2239, \hepth{0205217};
%
B.~C.~Vallilo,
\jhep{0212} (2002) 042, \hepth{0210064};
%
M.~J.~Chesterman, \jhep{0402} (2004) 011, \hepth{0212261};
%
I.~Oda, \plb{520} (2001) 398, \hepth{0302203};
B.~C.~Vallilo, 
 \jhep{0403} (2004) 037, \hepth{0307018};
M.~J.~Chesterman, \npb{703}, (2004) 400, \hepth{0404021};
O.~Chandia and B.~C.~Vallilo,
\jhep{0404} (2004) 041, \hepth{0401226};
%
S.~Guttenberg, J.~Knapp and M.~Kreuzer,
\jhep{0406} (2004) 030, \hepth{0405007};
%
  N.~Berkovits, \hepth{0409159};
%
T.~Oota,
\hepth{411036};
 N.~Berkovits, \hepth{0411170};
%
\bibitem{Berk0209ICTP}
  N.~Berkovits, 
  {\it ICTP Lectures on Covariant Quantization of the Superstring}, 
  \hepth{0209059}. 
\bibitem{stonybrook} 
  P.~A.~Grassi, G.~Policastro, M.~Porrati and P.~van Nieuwenhuizen, 
  \jhep{0210} (2002) 054, \hepth{0112162};
%
  P.~A.~Grassi, G.~Policastro and P.~van Nieuwenhuizen, 
  \jhep{0211} (2002) 001, \hepth{0202123};
  Adv. Theor. Math. Phys. {\bf 7} (2003) 499, \hepth{0206216};
  \plb{553} (2003) 96, \hepth{0209026};
  \hepth{0211095};
  Class. Quant. Grav. {\bf 20} (2003) S395, \hepth{0302147};
  \npb{676} (2004) 43, \hepth{0307056};
  \hepth{0402122};
%
P.~A.~Grassi and P.~van Nieuwenhuizen,
  \npb{702} (2004) 189, \hepth{0403209};
  \hepth{0408007};
P.~A.~Grassi and L.~Tamassia,
\jhep{0407} (2004) 071, \hepth{0405072};
L.~Anguelova, P.~A.~Grassi and P.~Vanhove,
\npb{702} (2004) 269, \hepth{0408171};
P.~A.~Grassi and P.~Vanhove,
\hepth{0411167}.
%
\bibitem{AK1}
Y.~Aisaka and Y.~Kazama, \jhep{0302} (2003) 017, \hepth{0212316}.
\bibitem{AK2}
Y.~Aisaka and Y.~Kazama, \jhep{0308} (2003) 047, \hepth{0305221}.
\bibitem{AK3}
Y.~Aisaka and Y.~Kazama, \jhep{0404} (2004) 070, \hepth{0404141}.
\bibitem{Oda0109} 
  I.~Oda and M.~Tonin, \plb{520} (2001) 398, hep-th/0109051;
  \plb{606} (2005) 218; \hepth{0409052}.
\bibitem{Matone0206} M.~Matone, L.~Mzzucato, I.~Oda, D.~Sorokin 
 and M.~Tonin, \npb{639} (2002) 182, \hepth{0206104}. 
\bibitem{BM0412}
N.~Berkovits and D.~Z.~Marchioro, \jhep{0501} (2005) 018, \hepth{0412198}. 
\bibitem{semiLC}
S.~Carlip, \npb{284} (1987) 365; 
R.~Kallosh and A.~Y.~Morozov,
Int.\ J.\ Mod.\ Phys.\ A {\bf 3} (1988) 1943; 
\bibitem{Siegel83}
W.~Siegel, \plb{128} (1983) 397.
\bibitem{BS}
L.~Brink and J.~H.~Schwarz, \plb{100} (1981) 310. 
\bibitem{Berkmembrane}
  N.~Berkovits, 
 \jhep{0209} (2002) 051, \hepth{0201151}.
\bibitem{Anguelovaetal}
L.~Anguelova, P.~A.~Grassi and P.~Vanhove,
\npb{702} (2004) 269, \hepth{0408171}.
%
%
\bibitem{HenTb} M.~Henneaux and C.~Teitelboim, 
  {\it Quantization of Gauge Systems}, Princeton University Press, 1992. 
%
\bibitem{Cartan} E.~Cartan, {\it Le\c{c}on sur la Theorie
 des Spineurs }, Hermann, Paris, 1937. \\
({\it The Theory of Spinors}, MIT Press, 1966. ) 
\bibitem{Cederwalletal}
M.~Cederwall, B.~E.~W.~Nilsson and D.~Tsimpis,
\jhep{0202} (2002) 009, \hepth{0110069}.
\end{thebibliography}
\end{document}